\documentclass[journal]{IEEEtran}

\usepackage[latin1]{inputenc}
\usepackage{amsmath}
\usepackage{graphicx}
\usepackage{listings}
\usepackage[dvipdfm,colorlinks,linkcolor=black,citecolor=black,urlcolor=black]{hyperref}
\usepackage{siunitx}
\usepackage{cite}
\usepackage{fixltx2e}
\usepackage{multirow}
\usepackage{threeparttable}

\begin{document}
\title{A Biological-Realtime Neuromorphic System in 28~nm CMOS using Low-Leakage Switched Capacitor Circuits}

\author{Christian Mayr$^1$, Johannes Partzsch$^2$, Marko Noack$^2$\\Stefan H\"anzsche$^2$, Stefan Scholze$^2$, Sebastian H\"oppner$^2$\\ Georg Ellguth$^2$, and Rene Sch\"uffny$^2$\\
$^1$ Institute of Neuroinformatics, University of Zurich and\\ ETH Zurich, Zurich, Switzerland\\
$^2$ Technische Universit\"at Dresden, Endowed Chair of Highly Parallel\\ VLSI Systems and Neuromorphic Circuits, Dresden, Germany,\\ email: cmayr@ini.uzh.ch}

\maketitle
\begin{abstract}
A switched-capacitor (SC) neuromorphic system for closed-loop neural coupling in 28~nm CMOS is presented, occupying 600~um by 600~um. It offers 128 input channels (i.e. presynaptic terminals), 8192 synapses and 64 output channels (i.e. neurons). Biologically realistic neuron and synapse dynamics are achieved via a faithful translation of the behavioural equations to SC circuits. As leakage currents significantly affect circuit behaviour at this technology node, dedicated compensation techniques are employed to achieve biological-realtime operation, with faithful reproduction of time constants of several 100~ms at room temperature. Power draw of the overall system is 1.9~mW.
\end{abstract}

{\bf Keywords:} biological-realtime neuromorphic system, switched capacitor neuromorphic circuits, biohybrid interface, deep submicron switched capacitor, low leakage switched capacitor

\section{Introduction}

There has been significant recent success in using neuromorphic circuits and/or neural network simulations in brain-machine interfaces. Examples include central pattern generators for spinal cord prostheses \cite{vogelstein08} or neural network based decoding filters for arm prostheses \cite{dethier13}. In order to achieve the millisecond to second time constants necessary for interfacing these neuromorphic circuits to biological circuits \cite{vogelstein08} or to realtime sensor/motor interfaces \cite{connor13,koenig02,mayr07d,mayr07c}, most analog implementations of neuromorphic circuits rely on so-called subthreshold circuits \cite{bartolozzi07}. 

 However, subthreshold circuits are hard to port to advanced CMOS techologies, since leakage currents rapidly increase with down-scaling, reaching the range of the desired signal currents. This is why even recent neuromorphic systems have been manufactured in quite old technologies \cite{indiveri10,moradi13}. Thus, with the exception of fully digital implementations \cite{galluppi12,seo12}, current neuromorphic systems are not able to participate in the technological advances and especially the system scaling offered by deep submicron processes.

These problems can be largely circumvented by using switched capacitor (SC) circuits \cite{vogelstein07,folowosele09b,noack14a}, which rely on charges and voltages to perform computation, not on currents. By replacing continuously flowing very small currents with their equivalent accumulated charge, equivalent signal levels are higher (and hence more controllable) and robust charge-based signal transmission and computation can be utilized.  

We present a neuromorphic system realized in SC circuit technique in a Super Low Power 28~nm CMOS technology, operating with a 1~V supply. The system is targeted at a closed loop interface to in-vitro cortical neuron cultures. This necessitates mimicking the memory and short-term decision making dynamics of the cortical network \cite{rolls13,mayr10a}, with timescales on the order of several 100~ms \cite{levy12,vogelstein08}. We implement the model of short term dynamics presented in \cite{noack12}, with transistor-level SC circuits derived from the high-level building blocks introduced in \cite{noack12}. The logical organisation of the synaptic matrix was adapted from \cite{noack10}.

Although the SC technique is inherently more robust to current and voltage noise than subthreshold circuits, on the timescales referred to above, stored charge signals can still be affected by the leakage currents of switching transistors. This effectively limits biological-realtime operation. Thus, we use a simplified version of the circuit techniques in \cite{ellguth06,ishida06} to reduce leakage currents to achieve longer time constants.  Compared to conventional biological interfacing solutions, \cite{chen13,lopez13}, no digital processing chain is necessary. The behavioural models that allow the system to couple into biological dynamics are directly implemented as discrete-time analog state circuitry, driven by incoming action potentials (i.e. spikes). At the same time, the SC approach makes the systems' behaviour widely digitally configurable. The use of 28nm CMOS eases integration with low-power digital systems.

The remainder of the paper is structured as follows. First, we introduce the overall system, followed by its digital and analog building blocks. We show how biologically realistic neuron and synapse behaviour as well as biological-realtime operation is achieved with SC CMOS circuit techniques. We then give detailed measurement results on the overall system and its individual components. Lastly, we discuss the significance of the results.

\section{Implementation}
\label{sec_implementation}
\subsection{Overall System}
\label{sec_design_system}

\begin{figure}
\centering
\includegraphics[width=0.5\textwidth]{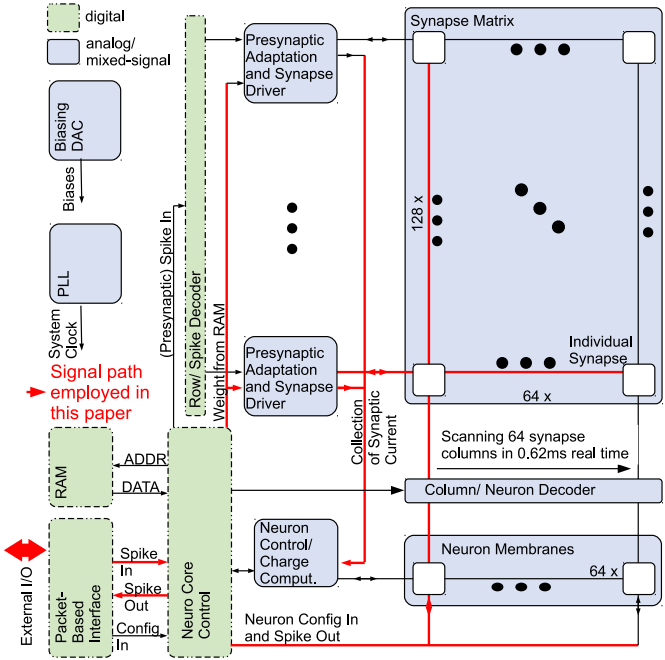}
\caption{\label{fig_overview}Overview of the neuromorphic system including the synaptic matrix and the other neuromorphic mixed signal SC blocks, digital control, synaptic weight RAM, biasing digital-analog converter (DAC), phase-locked loop (PLL) clock input and serial packet input/output (I/O).}
\end{figure}

Fig. \ref{fig_overview} gives an overview of the system. 128 input circuits on the left side implement presynaptic short term dynamics for their respective row in the synaptic matrix \cite{noack12}, while the 64 Leaky Integrate and Fire (LIAF) neurons shown on the bottom are driven by their respective column, providing the output (i.e. stimulation) signal as a function of the 8192 synapses in the matrix coupling presynaptic inputs to the neurons. Synaptic weights are stored in a RAM block on the side of the matrix. 

The entire driving circuitry of presynapses, synapses and neurons is situated on the left hand side of the matrix. In real-time operation, a state machine cycles through the columns of the synaptic matrix in 0.62~ms. At the start of the cycle, the input pulses that were registered during the last cycle are forwarded to the driver circuits and the corresponding presynaptic adaptation state is computed. Then, each synaptic column and its corresponding output neuron is activated sequentially, weighting the presynaptic pulses by the corresponding synapse state and synaptic weight, integrating them on the postsynaptic neuron and applying the leaky decay term to the neuron. Details on the cycle process can be found in Sec. \ref{sec_design_neuron}. In effect, the switched capacitor neuron and synapse matrix behaves as a fixed-timestep neural simulator with a 0.62~ms time resolution, with neuron and synapse states stored in the matrix and updates to the states carried out via the active driver circuits on the left side of the matrix \cite{noack10,mayr10b}. 

\begin{figure}
\centering
\includegraphics[width=0.5\textwidth]{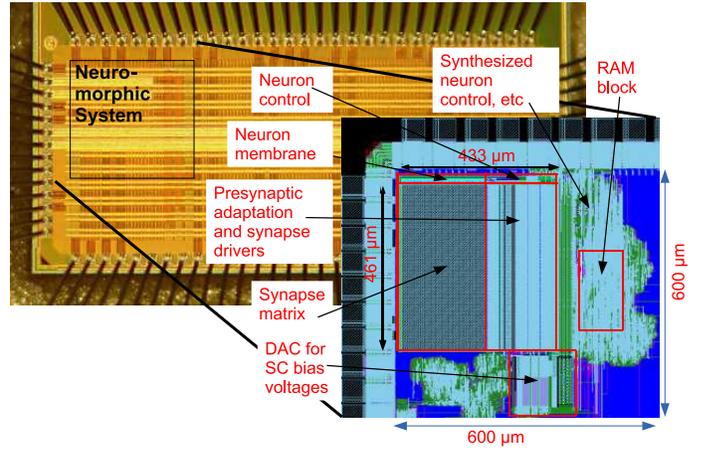}
\caption{\label{fig_die}28~nm die picture with location of neuromorphic system outlined and the corresponding layout. The overall IC is 1.5~mm by 3~mm, with various test structures in addition to the neuromorphic system. The layout view shows detailed placement of the single building blocks of the neuromorphic system (see Fig. 1).}
\end{figure}

A picture of the manufactured IC is shown in Fig. \ref{fig_die}. The circuit design utilizes core devices of the SLP 28~nm technology only. In contrast to the current biasing usually employed in neuromorphic ICs \cite{yang12}, the neuromorphic SC circuits are governed by voltages for amplitude settings and digital configuration for time constants. Correspondingly, there is a multi-output R-ladder based digital to analog converter (DAC) situated below the matrix in Fig. \ref{fig_die}. It provides the bias voltages for e.g. postsynaptic current (PSC) scaling, neuron thresholds or reset voltages. To reduce the area of the DAC, neuromorphic elements have been assigned to groups sharing the same bias voltages. Group size is 16, so that neurons 0 to 15 share their biases, and synapse drivers and presynaptic adaptation 0 to 15 also share their biases, etc.

Time constants are set via counters that govern the switching cycles of the SC circuits. Thus, scaling of the clock frequency effectively scales the speed of the system, keeping the 0.62~ms resolution relative to the chosen time base. 
The neuromorphic system was designed for speeds from biological realtime up to an acceleration of 100.
As the time constants scale with the clock and the DAC amplitude settings are independent of clock speed, the same configuration for all parameters can be used irrespective of the speed-up, nominally giving the same results.

Communication with the system is provided by a joint test action group (JTAG) interface, implementing a generic packet-based protocol for both pulse and configuration data. Additionally, two configurable test outputs allow for monitoring analog voltages, such as membrane potentials.
With its minimal interface, using only six signal pins and two bias pins (one bias current and one pin for common mode voltage), the neuromorphic system can be easily integrated into a multi-core system.

\subsection{Digital System Design}
\label{sec_design_digital}

Similar to the communication setup in \cite{hartmann10,scholze11a}, the neuromorphic system employs a unified packet-based interface for configuration and incoming/outgoing pulse data.
Data exchange is realized via an input first-in-first-out (FIFO) buffer and an output FIFO buffer.
For system integration, only a write and read interface to these FIFOs has to be provided, which is done via JTAG in the current implementation.
Each data packet has a 32~bit payload and a 16~bit header.
For input data, the header includes 4~bits of type and 12~bits of address information.
For output data, the header only contains a 5~bit type identifier.

Input spikes are sent to the neuromorphic system as addresses of 7~bits and one enable bit, so that four spikes fit in one data packet.
Output spikes are collected over one matrix cycle and stored as one bit per neuron. 
If at least one neuron spiked in the current cycle, the 64~bit spike vector over all neurons is sent to the output FIFO, forming two separate entries.
Similar to the grouping for the analog parameters, the digital parameters are also shared among groups of 16 each for presynaptic circuits and neurons, which reduces the digital configuration space.

All digital components and the SC circuits are clocked by an on-chip phase-locked loop (PLL) \cite{hoeppner13}. It produces an internal fixed frequency of 2~GHz that is downscaled to a 330~MHz output. The neuromorphic system employs an 8~bit configurable clock divider that  allows for further downscaling of the clock frequency. Biological-realtime operation corresponds to a divider value of 100, i.e. a clock frequency of 3.3~MHz for the state machine of the neuromorphic system and a matrix update cycle of \SI{0.62}{\milli\second}, as mentioned in Sec. \ref{sec_design_system}. For the maximum speed-up factor of 100, the divider value is 1, resulting in a clock frequency of 330~MHz and a matrix update cycle of \SI{6.2}{\micro\second}. With respect to the update frequency of the matrix, the clock is somewhat high, which is due to the fact that non-overlapping switching signals for the SC-components are derived from it, see also the signal edges in Fig.  \ref{fig_circuit_timing}. The maximum speed-up factor is partially limited by the high clock frequency needed for the digital components, but the actual limit is due to the RC time constants of the SC circuits, as explained later.

\subsection{Presynaptic Adaptation and Synaptic Long-Term Plasticity}
\label{sec_design_presynapse}

\begin{figure}
\centering
\includegraphics[width=0.48\textwidth]{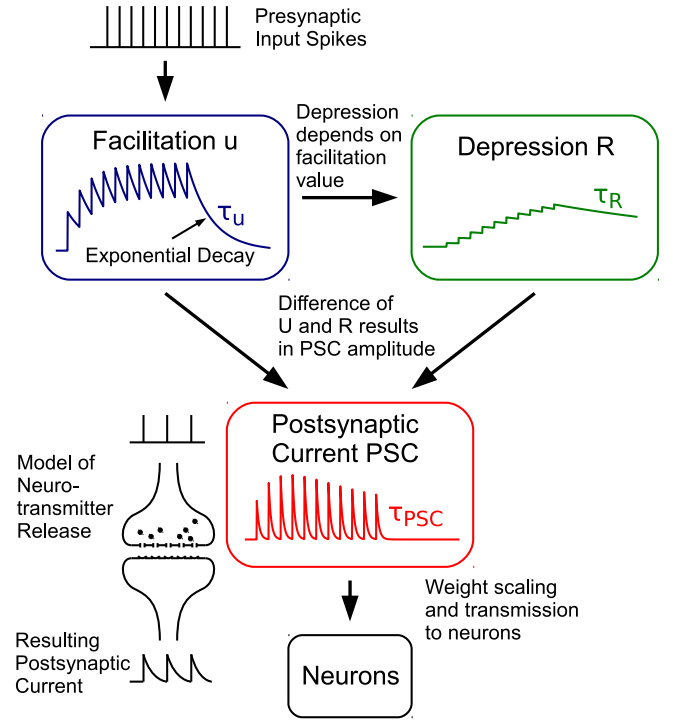}
\caption{\label{fig_stp}Overview of the presynaptic adaptation circuit. The combination of facilitation and depression mechanisms modulates the amplitude of the PSC traces. Each subcircuit holds the corresponding model variable which is updated at incoming presynaptic spikes and decays between spikes.}
\end{figure}

The presynaptic adaptation circuit (see Fig. \ref{fig_stp}) implements the model of synaptic dynamics proposed in \cite{noack12}, which is derived from biological measurements \cite{markram98}. It is capable of reproducing depression, facilitation and combinations of both mechanisms. The circuit produces an output voltage $V_{psc}$, which represents the waveform of exponentially decaying PSCs:

\begin{equation}
V_{psc}(t)=\hat{V}_{psc,n}\cdot\mathrm{exp}(-\frac{t}{\tau_{psc}})\;,
\label{eq_psc}
\end{equation}
where $\hat{V}_{psc,n}$ is the amplitude of the n-th PSC. Since the short-term adapation circuitry makes use of SC circuits, the resulting PSC voltage trace is time discrete. The time constant $\tau_{psc}$ of the PSC decay, as well as the time constants for depression $\tau_R$ and facilitation $\tau_u$, can be adjusted. The impact of the facilitation and depression mechanisms can be controlled by the digital parameters $U$ and $\alpha$, respectively. For details, please refer to \cite{noack12}. For a list of configurable parameters and their tuning range see Tab. \ref{tab_parameters}.

\begin{table}
\caption{\label{tab_parameters}List of parameters for short-term adaptation and neuron circuit.}
\centering
\begin{threeparttable}
        \begin{tabular}{|l|l|l|l|}
            \hline
            Parameter & Description & Range\tnote{1} \\
            \hline
			$U$ & Utilization of syn. efficacy \cite{markram98,noack12} & \numrange[range-phrase = -- ]{0}{0.98} \\
			$\alpha$ & Strength of depression \cite{noack12} & \numrange[range-phrase = -- ]{0}{0.98} \\
			$\tau_u$ & Facilitation time constant & \SIrange[range-units=single,range-phrase = -- ]{9.6}{605}{\milli\second}, inf.\\
			$\tau_R$ & Depression time constant & \SIrange[range-units=single,range-phrase = -- ]{9.6}{605}{\milli\second}, inf.\\
			$\tau_{PSC}$ & PSC time constant & \SIrange[range-units=single,range-phrase = -- ]{1.2}{74.5}{\milli\second}, inf.\\
        	$V_{reset}$ & Reset potential of LIAF Neuron & -250 to 250~mV\\
        	$V_{thresh}$ & Firing threshold of LIAF neuron & -250 to 250~mV\\
        	$\tau_m$ & Membrane time constant & \SIrange[range-units=single,range-phrase = -- ]{1.2}{74.5}{\milli\second}, inf.\\
        	$S$ & Speed-up factor & \numrange[range-phrase = -- ]{1}{100}\\
        	\hline
        \end{tabular}
        \begin{tablenotes}
\item[1] Voltages are set digitally with 7~bit precision using digital-to-analog converters. Time constants are configurable via 6~bit counter registers. All time constants are given for biological-realtime operation and can be scaled according to the speed-up factor $S$, e.g. $\tilde\tau_u=\tau_u/S$. The resting voltage of the LIAF neuron is fixed at \SI{0}{\volt}.
\end{tablenotes}
\end{threeparttable}
\end{table}

\begin{figure}
\centering
\includegraphics[width=0.35\textwidth]{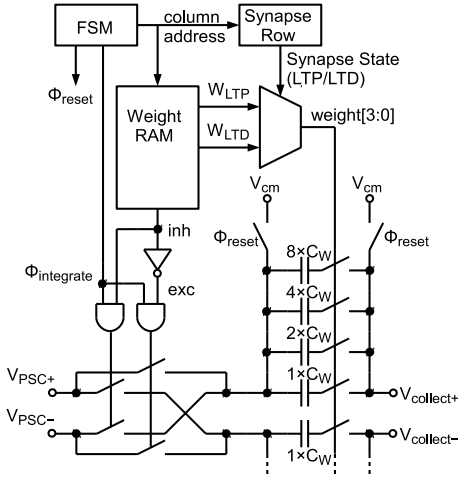}
\caption{\label{fig_circuit_weight_scaling}Weight scaling circuit with digital control and binary-weighted capacitors.}
\end{figure}

The long term plasticity model chosen for this neuromorphic system is the stochastic stop learning synapse of \cite{brader07}. It is based on modifying the synaptic state as a function of the presynaptic spike and the postsynaptic membrane voltage. In our implementation of the rule, when a column gets activated during the matrix cycle, the analog synaptic state held in the corresponding synaptic capacitance in the synaptic matrix is read out. It is then modified according to the equations in \cite{brader07}, with appropriate configurable parameters. As this paper focuses on the overall neuromorphic system and its static operation, the reader is referred to the companion paper \cite{noack14b} for an in-depth circuit description and detailed measurements of long- and short term plasticity in this neuromorphic system.

Based on the synapse state, the PSC amplitude is then modified by a weight scaling circuit. There is one of these per synapse row, located in the presynaptic adaptation circuit. As can be seen in Fig. \ref{fig_circuit_weight_scaling}, all synapses are addressed sequentially by a state machine (see Fig. \ref{fig_circuit_timing} for the timing diagram, which also shows the entire matrix cycle in relation to the synapse and neuron driver signals). Corresponding to the synapse address, 4~bit long-term potentation (LTP) and long-term depression (LTD) weight values $W_{LTP}$ and $W_{LTD}$ are read from a RAM. To scale the presynaptically computed PSC by the long term plasticity, the synapse state is collapsed into a binary state value, which can be either potentiated or depressed \cite{brader07}. Depending on this synapse state, the 4~bit LTP or LTD weight is then selected by a multiplexer. The switches at the four binary-weighted capacitors are closed according to the given weight value. After selecting a synapse, the weight capacitors C$_\mathrm{W}$ are initially reset to \SI{0}{\volt}. In the following integration phase the differential PSC voltage $V_{psc}$ is applied to the input of the weight scaling circuit. The charge is then transmitted by the capacitors to the neuron circuit (see Sec. \ref{sec_design_neuron}). Additionally, the weight scaling circuit offers the possibility to configure the synapses as either inhibitory or excitatory. This configuration bit is also stored in RAM.

While presynaptic drivers for almost all synapses are activated by an incoming pulse, synapse row 127 is always active, with a constant charge. This charge can be modulated indvidually for each neuron by setting the synaptic weight of row 127 and the column corresponding to the neuron. This way, a constant background current with a 4bit weight and inhibitory or excitatory effect can be set.

\subsection{Switched Capacitor Neuron}
\label{sec_design_neuron}

\begin{figure}
\centering
\includegraphics[width=0.48\textwidth]{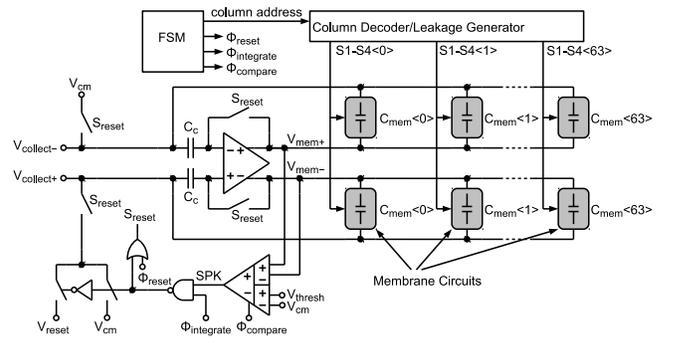}
\caption{\label{fig_circuit_neuron}Neuron circuit. A detailed diagram of the membrane circuit can be found in Fig. \ref{fig_lowleakage_switch}c.}
\end{figure}

\begin{figure}
\centering
\includegraphics[width=0.48\textwidth]{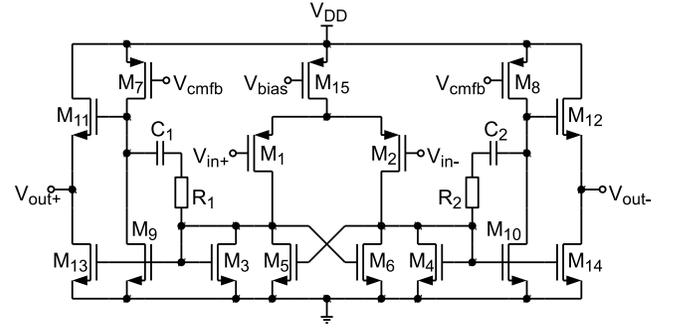}
\caption{\label{fig_circuit_opamp}Fully-differential opamp used in the neuron circuit.}
\end{figure}

The neuron circuit implements an LIAF neuron model:
\begin{equation}
\frac{\mathrm{d}V_{mem}}{\mathrm{d}t}=-\frac{V_{mem}}{\tau_m} + \frac{I_{syn}}{C_{mem}}\,,
\end{equation}
with membrane potential $V_{mem}$, membrane capacitance $C_{mem}$ and membrane time constant $\tau_m$. $I_{syn}$ is the sum of all PSCs. If $V_{mem}$ reaches the firing threshold $V_{thresh}$ a spike is emitted and the membrane is reset to $V_{reset}$. The parameters of the LIAF neuron and their tuning ranges are listed in Tab. \ref{tab_parameters}. As can be seen in Fig. \ref{fig_circuit_neuron}, the 64 fully-differential membrane circuits are located on one row and share one driver circuit. The membrane circuits are sequentially switched to active and the PSC output of all 128 weight scaling circuits (Fig. \ref{fig_circuit_weight_scaling}) are summed on node $V_{collect}$ as a charge. The charge on the global summing node $V_{collect}$ is integrated on the currently selected membrane capacitance by the driver circuit, which is basically an SC integrator.

The integrator's opamp circuit is shown in Fig. \ref{fig_circuit_opamp}. A two-stage architecture has been chosen to overcome the difficulties of stacking transistors at very low supply voltages. In order to enhance the opamp's gain, a boosting technique has been applied \cite{dessouky00}, where the load of the first stage has been split into cross-coupled transistors, providing partial positive feedback. Stability is derived by Miller compensation and the common-mode voltage of the output stage is controlled by an SC common-mode feedback circuit. Slew rate performance is enhanced by additional source followers at the output, which is required at high speed-up factors. The bias current scales well with the speed-up, so that the opamp consumes \SI{300}{\nano\watt} at biological realtime and \SI{30}{\micro\watt} at a speed-up of 100.

\begin{figure}
\centering
\includegraphics[width=0.48\textwidth]{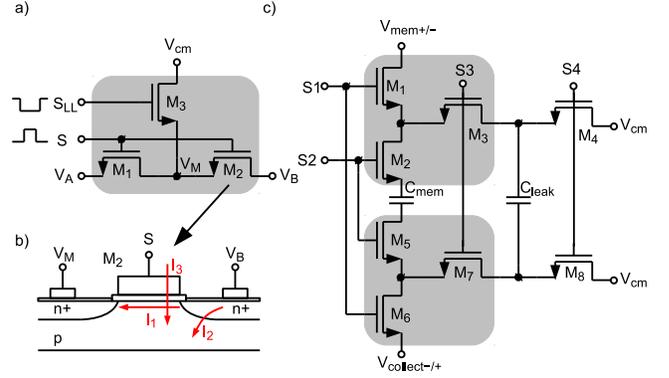}
\caption{\label{fig_lowleakage_switch}(a) Low-leakage switch configuration. (b) Cross-section of MOS Transistor M2 with denoted subthreshold leakage ($I_1$), junction leakage ($I_2$) and gate leakage ($I_3$). (c) Membrane circuit with low-leakage switches (grey boxes) and SC leakage generation.}
\end{figure}

For biological-realtime operation, large membrane time constants in the order of \SI{100}{\milli\second} are required. Since leakage currents heavily increase when scaling technologies down below \SI{100}{\nano\meter} \cite{roy03}, a dedicated low-leakage switch similar to those in \cite{ellguth06} and \cite{ishida06} has been used (see Fig. \ref{fig_lowleakage_switch}a), which operates as follows. If the membrane circuit is inactive the membrane capacitance is fully decoupled from the rest of the circuit by turning off M1 and M2. The middle node $V_M$ of the T-switch is set to the common-mode voltage $V_{cm}$. This reduces the drain-source voltage over M2, which in turn reduces the subthreshold current flowing through the channel (see $I_1$ in Fig. \ref{fig_lowleakage_switch}b). In order to decrease junction leakage ($I_2$), minimally sized source/drain areas have been used. While this sizing aids in real-time operation of the matrix, it also defines the upper limit of the speed-up compared to biological realtime (i.e. the factor 100 mentioned in Sec. \ref{sec_design_system}), as the switch resistance determines the RC time constant of the SC circuits, limiting the charge transfer speed.

A further advantage of the low-leakage switch is that the decoupling via the middle node makes the leakage currents independent of the opamp output ($V_{mem+}$,$V_{mem-}$). Gate leakage ($I_3$) has no impact in the off-state of the switch and simulations have shown that the effect in the short on-states is negligible. In contrast to the complimentary transmission gates of \cite{ellguth06,ishida06}, the voltage range chosen here allows to use NMOS devices only, reducing leakage currents and circuit complexity. Regular V$_\mathrm{th}$ transistors from the core library were used, rather than dedicated low-V$_\mathrm{th}$ devices as in \cite{ishida06}. Compared to \cite{ellguth06}, the middle node is held at $V_{cm}$ to reduce $V_{DS}$-caused channel leakage. The presynaptic adaptation circuit of Sec. \ref{sec_design_presynapse} also employs this low leakage switch to achieve its time constants.

\begin{figure}
\centering
\includegraphics[width=0.48\textwidth]{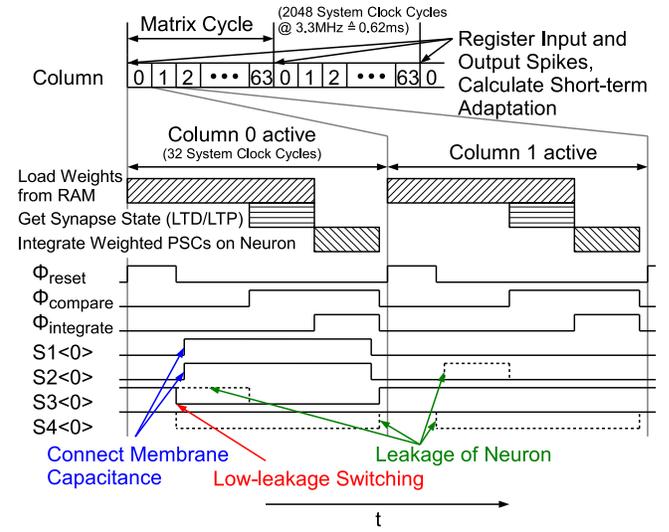}
\caption{\label{fig_circuit_timing}Timing diagram of the matrix cycle (top), sequence of loading synaptic weights and integration on the neuron membrane (middle) and timing diagram of switching phases and control signals for the membrane circuits (bottom). The dashed lines indicate whether switches are on or off when a leakage event occurs. Shifted clock edges denote digitally-generated non-overlapping switch signals required by the SC circuits.}
\end{figure}

While the circuit combats undesired leakage currents, it is also used to implement an intentional, configurable leakage mechanism to complete the LIAF neuron model. This is directly implemented in the individual membrane circuits (see Fig. \ref{fig_lowleakage_switch}c). A small capacitance $C_{leak}=\SI{5}{\femto\farad}$ is discharged and then shunted to the membrane capacitance $C_{mem}=\SI{75}{\femto\farad}$, leading to a charge equalization. This process is triggered periodically and thus lets the membrane voltage decay exponentially towards \SI{0}{\volt} differential voltage when no synaptic input is applied. The membrane time constant $\tau_m$ is controlled by the switching frequency, alternatively expressed as the period $T_{leak}$ between leakage events:
\begin{equation}
T_{leak} = -\tau_m \cdot\mathrm{ln}(\frac{C_{mem}}{C_{mem}+C_{leak}})\,.
\end{equation}
Since $T_{leak}$ is derived from the system clock, the membrane time constant and all other time constants generated by SC circuits on the chip are proportional to the speed-up factor.
Fig. \ref{fig_circuit_timing} shows the four control signals used by the membrane circuit of column 0.

In order to avoid a permanent integration of the opamp's offset voltage generated by device mismatch, an offset compensation technique has been applied \cite{enz96}. In the reset phase $\Phi_{reset}$, unity gain feedback is applied to the opamp. Thus, the output offset voltage is visible at the input. Since $V_{collect}$ is reset to $V_{cm}$ at this time, the opamp offset is sampled on the compensation capacitance $C_c$. In the following integration phase $\Phi_{integrate}$, i.e. the phase where the PSC charges are transferred to the membrane, the opamp offset is substracted from the input voltage. 

In the comparison phase $\Phi_{compare}$, the membrane voltage is compared against the firing threshold. The comparator circuit consists of an offset compensated preamplifier and a dynamic latch. If $V_{mem+}-V_{mem-} > V_{thresh}-V_{cm}$, a spike is detected and the membrane voltage is reset. Due to the single-ended nature of the biasing voltages $V_{reset}$ and $V_{thresh}$, the reset is done in an asymmetric fashion, but is compensated by the opamp's common-mode feedback. 

\section{Results}
\label{sec_results}

As detailed in Sec. \ref{sec_design_digital}, the entire system is ratiometric with respect to the clock frequency. That is, the system clock can be scaled so that the system operates anywhere from biological realtime up to a factor of 100 faster. Realtime operation was used for the measurements in this paper, as the effectiveness of the leakage current techniques becomes most evident there. In addition, operation in biological realtime is the most interesting regime in terms of computation, as it allows interfacing with e.g. neuromorphic image sensors in real time. The IC and its board are operated at ambient temperature, i.e. no special measures are undertaken to cool the IC.

\subsection{Measurement of the Presynaptic Adaptation}

For measuring the presynaptic adaptation circuits, the two analog test outputs were captured using an oscilloscope, allowing the simultaneous measurement of the PSC voltage of the first presynaptic circuit and the membrane voltage of one neuron.
The aquired data was averaged over time bins of 0.1-0.3ms to reduce the effect of noise.

\begin{figure}
\centering
\includegraphics[width=0.5\textwidth]{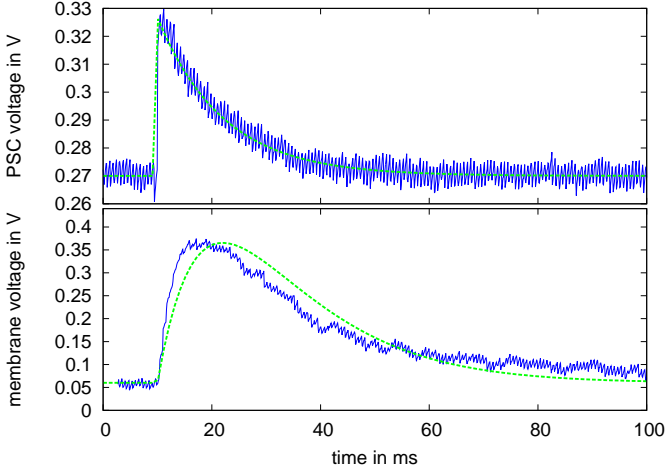}
\caption{\label{fig_wave_psp} Measurement (blue curves) of the PSC (top) and PSP (bottom) waveforms for parameters $\tau_\mathrm{PSC}=\tau_\mathrm{mem}=12$~ms. The nominal curves with the same time constants and fitted amplitude are shown as dashed green lines: PSC curve following Eq. \ref{eq_psc}, PSP according to $\alpha$-shape with $V(t)\sim t/\tau\cdot \mathrm{exp}(-t/\tau)$.}
\end{figure}

Figure \ref{fig_wave_psp} shows an example of a single postsynaptic potential (PSP).
Compared to the expected $\alpha$-shaped curve, the measurement shows a slightly sharper onset, indicating a mismatch in the actual time constants from the nominal values.
The corresponding PSC waveform matches with the nominal time constant well (see upper plot in Fig. \ref{fig_wave_psp}).
Thus, the mismatch can be attributed to a mismatch in the membrane leakage.
As the leakage mechanisms and capacitance sizes are the same in both cases, we attribute the additional leakage in the membrane to the 128 connected PSC outputs.

\begin{figure}
\centering
\includegraphics[width=0.5\textwidth]{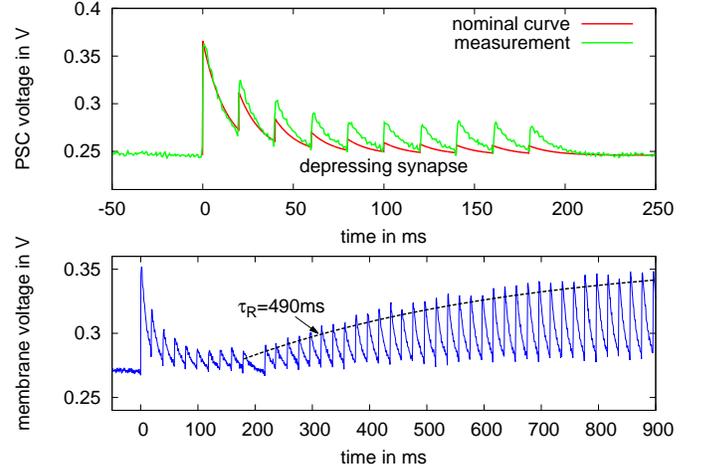}
\caption{\label{fig_presyn_depression} PSC voltage traces of a depressing synapse with parameters: $\tau_u=10$~ms, $\tau_R=490$~ms, $\tau_\mathrm{PSC}=13$~ms, $\tau_\mathrm{mem}=1.2$~ms, $U=0.96$, $\alpha=0.5$. Top: Synapse stimulated with 10 spikes at 50~Hz rate (green). The nominal time course for the PSC voltage with these parameters and fitted offset is drawn in red. Bottom: Same initial stimulation, but adaptation switched off after 10 spikes ($\alpha=0$), so that the synapse relaxes with depression time constant $\tau_R$. Nominal decay with $\tau_R=490$~ms drawn as a dashed line.}
\end{figure}

\begin{figure}
\centering
\includegraphics[width=0.5\textwidth]{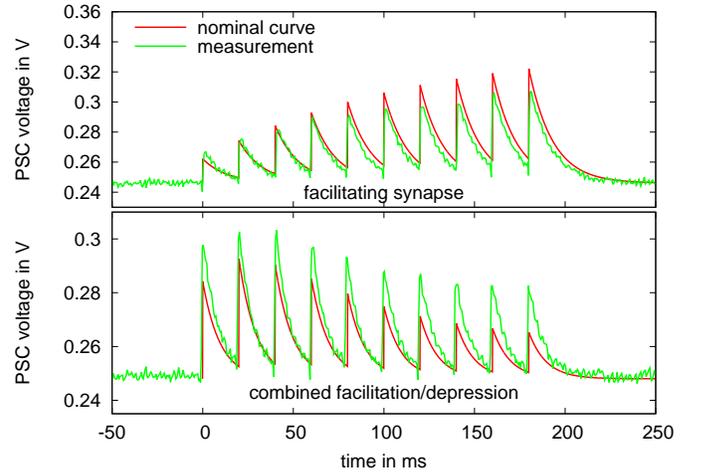}
\caption{\label{fig_presyn_facil_comb} Plot of measured PSC waveforms (green). Top: Facilitating synapse with parameters: $\tau_u=490$ms, $\tau_R=10$ms, $\tau_\mathrm{PSC}=13$ms, $U=0.13$, $\alpha=0.86$. Bottom: Simultaneously acting facilitation and depression; parameters: $\tau_u=300$ms, $\tau_R=300$ms, $\tau_\mathrm{PSC}=10$ms, $U=0.29$, $\alpha=0.5$. Same stimulation as in Fig. \ref{fig_presyn_depression} (10 spikes at 50Hz). PSC curves with nominal parameters and fitted offset are shown in red.}
\end{figure}

To evaluate the presynaptic adaptation performance, we stimulated a presynaptic circuit with a regular spike train, choosing various parameter settings to mimic different adaptation types.
Results are shown in Figs. \ref{fig_presyn_depression} and \ref{fig_presyn_facil_comb}.
The measurements agree well with the nominal time courses even without calibrating any parameters (note that for the nominal curves, only the offset of the read-out amplifier was fitted).
They differ mainly in the adaptation strength, i.e. in the ratio between highest and lowest PSC amplitude, which is smaller in the measured curves.
This effect is most prominent for the depressing synapse.
Also, for the synapse with combined facilitation and depression, the total amplitude is maybe 20\% too small, see lower half of Fig. \ref{fig_presyn_facil_comb}.
These effects may be caused by charge injection effects, resulting in voltage offsets during updates of the adaptation variables at incoming spikes.
Part of the effects may also be explained by the effective time constants being too small.
To distinguish between these two effects, we measured the relaxation of a depressed synapse when adaptation was turned off, see lower plot in Fig. \ref{fig_presyn_depression}.
The PSC amplitudes should progress according to an exponential with the depression time constant $\tau_R$ in this case.
This resembles the measurements well.
From this, we infer that the mismatch seen for the depressing synapse is mainly caused by deviating update amplitudes of the depression variable.

Overall, the measurement results show that leakage, charge injection and capacitance mismatch only have a minor impact on the time course of the state variables, showing faithful reproduction of time constants on the order of several hundred milliseconds.

\subsection{Characterization of the LIAF Neuron}
\label{sec_liaf_neuron}

We measured the transfer functions of the LIAF neurons in order to characterize variations between neurons and performance of the leakage circuit.
Single neurons were stimulated with regular spike trains at different rates and their output rates were measured over a period of 10 seconds.

\begin{figure}
\centering
\includegraphics[width=0.5\textwidth]{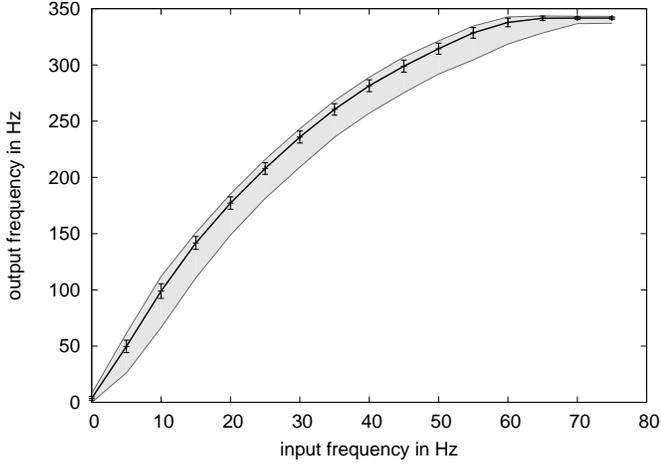}
\caption{\label{fig_liaf_transfer_neuron} Transfer function over all neurons, with the leaky term of the neurons switched off. The input rate is applied in parallel to 5 synapses of a neuron. Error bars denote the standard deviation over all 64 neurons. The gray area shows the range of all transfer functions}
\end{figure}

Fig. \ref{fig_liaf_transfer_neuron} shows results for all 64 neurons of one chip with leakage switched off.
As expected, the curve increases linearly at low rates, while saturating at high rates, which is caused by saturation of the PSC voltage of the presynaptic circuits.
The overall variation between neurons is quite low.
A few neurons generally exhibit a lower output frequency.
This is especially the case for neuron 0, which may be affected by additional parasitic capacitance at the border of the synaptic matrix.

\begin{figure}
\centering
\includegraphics[width=0.5\textwidth]{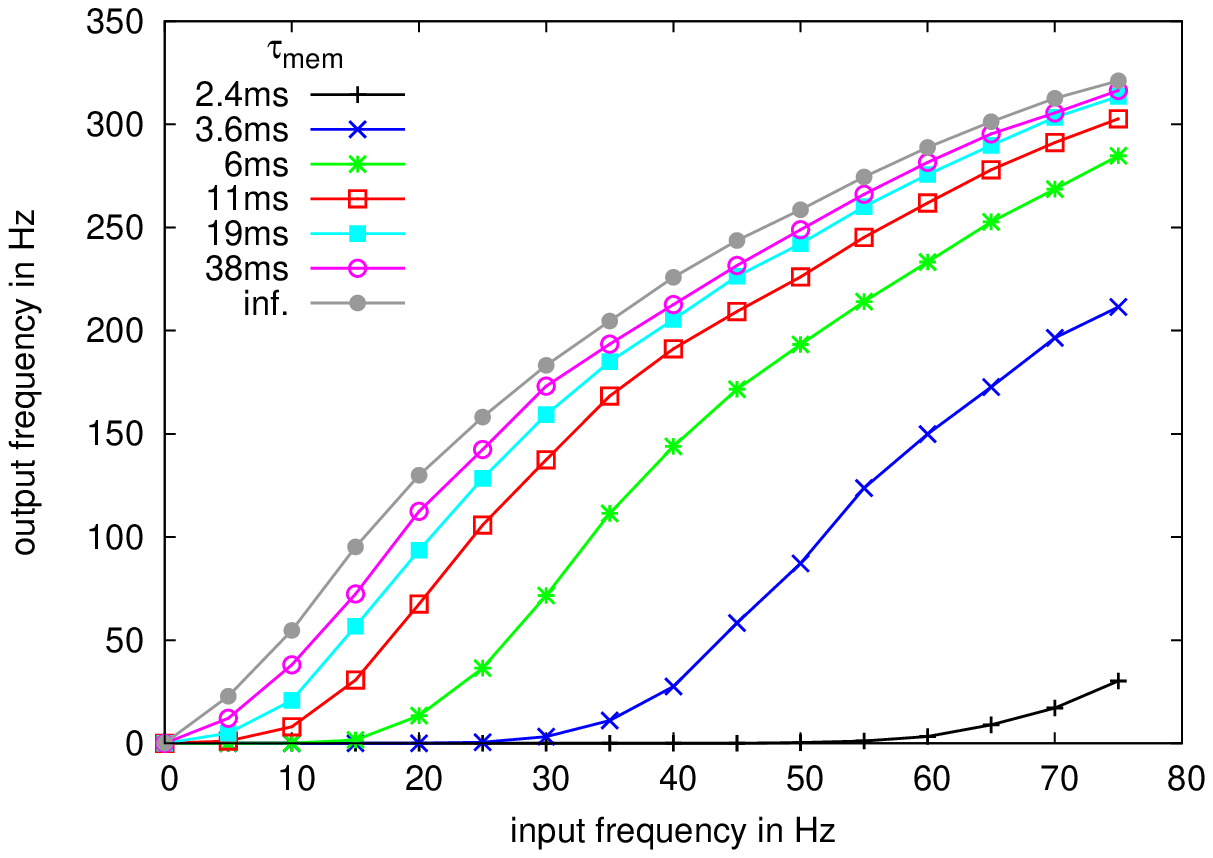}\\
\includegraphics[width=0.5\textwidth]{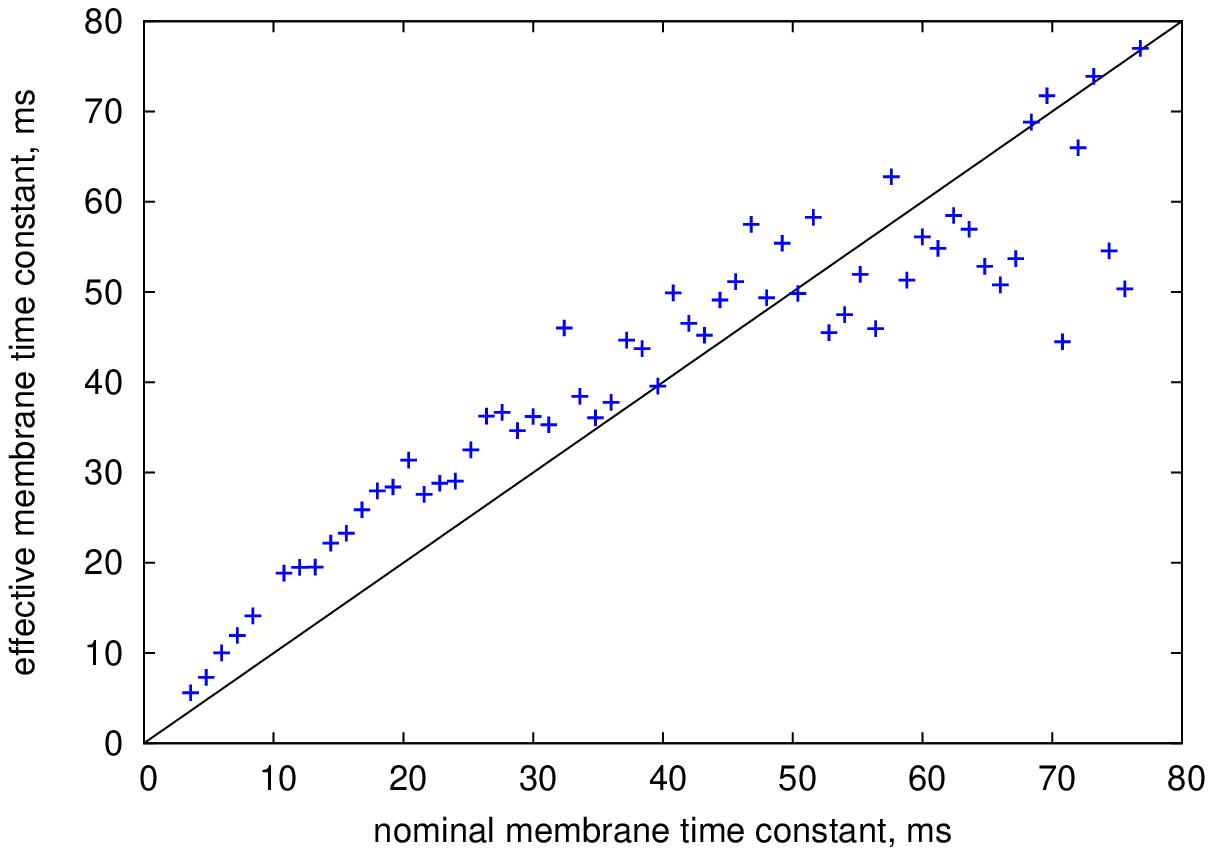}
\caption{\label{fig_liaf_transfer_taumem} Top: Transfer function of one neuron for different settings of the membrane time constant. The input rate is applied in parallel to 5 synapses of the neuron. Bottom: Neuron time constant $\tau_\mathrm{mem}$ extracted from the onset frequency $f_\mathrm{on}$ of the transfer function for the different time constant settings. The proportionality factor between time constant and onset frequency was chosen as $\tau_\mathrm{mem}\cdot f_\mathrm{on}=200\mathrm{ms}\cdot \mathrm{Hz}$ to fit the results.}
\end{figure}

Fig. \ref{fig_liaf_transfer_taumem} summarizes measurements for different membrane time constants.
As shown in the upper graph, the onset of the transfer functions varies with the time constant setting, as expected for a LIAF neuron.
Ideally, the onset frequency should be inversely proportional to the membrane time constant.
We used this relationship to compare the effective time constant with the configured settings.
The onset frequency was determined for each transfer function by performing a linear fit in the output frequency range of 50~Hz to 150~Hz, not taking onset and saturation effects into account.
Results are shown in the lower graph of Fig. \ref{fig_liaf_transfer_taumem}.
Note that this method is less accurate at larger time constants, where the onset frequency is close to zero, so that small absolute deviations in frequency result in high deviations in the final result. The effective time constants follow the nominal setting linearly for low values, while the slope of the curve decreases at higher values.
This effect may be caused by leakage, but may as well be due to a systematic offset in the transfer functions.

\subsection{Characterization of the Synaptic Transfer Function}

\begin{figure}
\centering
\includegraphics[width=0.5\textwidth]{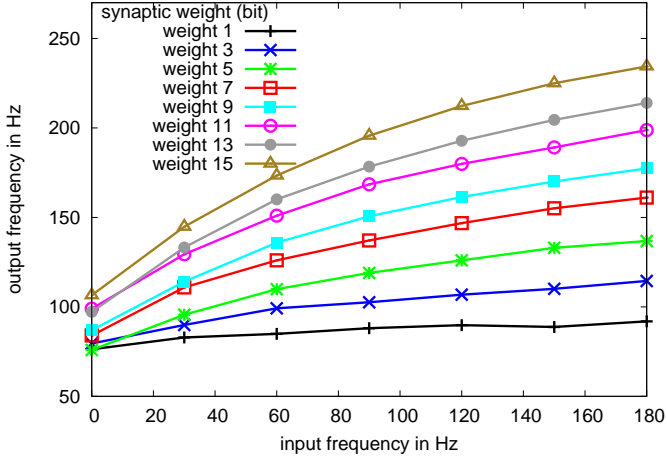}
\caption{\label{fig_synapse_sweep} Transfer function of one neuron for different synaptic weight settings, input stimulus applied to one synapse.}
\end{figure}

In order to characterize the synaptic transfer function, a fixed-rate pulse train is applied to a single synapse and the resulting firing rate of the postsynaptic neuron is measured. The neuron is configured for integrate-and-fire behaviour (with $\tau_\mathrm{mem}$ set to infinity) to achieve a linear relation between input and output firing rate. As can be seen in Fig. \ref{fig_synapse_sweep}, the individual curves show a smooth progression in output firing rate for an increase in input rate. Due to the PSC saturation effect mentioned in Sec. \ref{sec_liaf_neuron}, the relation between input and output firing rate declines to below linear for high input rates.

As can be seen from the curve intercept on the output frequency axis, a constant background current is applied to the neuron (via synapse row 127, compare Sec. \ref{sec_design_presynapse}) that sets its unstimulated firing rate at circa 80-105~Hz. From Fig. \ref{fig_liaf_transfer_taumem}, it can be seen that the neuron reacts well to very low rates of synaptic input even without background current. However, if the neuron intrinsically fires at a low rate for low input firing rates, charge injection and other small-signal detrimental effects partially mask the effect of a synaptic weight increase. Thus, this background current is applied to set the intrinsic neuron firing to a high rate, enabling the analysis of the synaptic weight on a 4~bit resolution level.   

\begin{figure}
\centering
\includegraphics[width=0.5\textwidth]{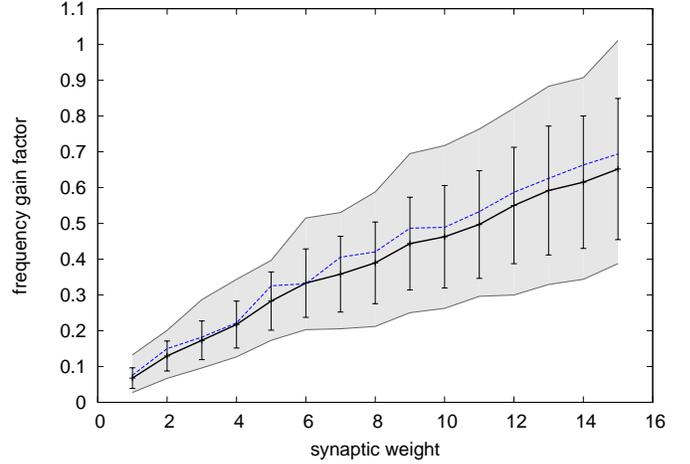}
\caption{\label{fig_synapse_transfer} Frequency gain factor (i.e. slopes of the curves in Fig. \ref{fig_synapse_sweep}) of neuron transfer functions over synaptic weight setting. The blue, dashed curve shows a sample curve of a sweep of one synapse as derived from the data in Fig. \ref{fig_synapse_sweep}. The black curve represents the mean of 20 randomly chosen synapses of one neuron. Error bars denote the standard deviation over the 20 synapses, the gray area shows the range of all weight transfer functions.}
\end{figure}

When sweeping the synaptic weight in Fig. \ref{fig_synapse_sweep}, the curves exhibit a linear progression in slope, showing the 4~bit accuracy of the synaptic weight scaling capacitances in Fig. \ref{fig_circuit_weight_scaling}. For the plots in Fig. \ref{fig_synapse_transfer}, the slopes of the curves in Fig. \ref{fig_synapse_sweep} are derived by fitting a linear function to the data points from 0 to 100~Hz input frequency. The blue dashed line shows the slopes derived for the synapse in Fig. \ref{fig_synapse_sweep}. This weight sweep was carried out for 20 synapses of one neuron on a single chip. As can be seen from the sample curve, the slope progression across synaptic weights is actually far better behaved than could be implied by the error bars in Fig. \ref{fig_synapse_transfer}. The large spread of curves is mainly due to the scaling error of V$_\mathrm{PSC}$ (compare Fig. \ref{fig_circuit_weight_scaling}). This error tends to even out when using several PSC circuits, as for the measurements in Sec. \ref{sec_liaf_neuron}, which use several PSC inputs and thus do not show such a large spread. V$_\mathrm{PSC}$ can also be calibrated to some extent via the individual DAC settings. However, this was not carried out for the above characterization, as we wanted to obtain an estimate for the typical spread that can be expected on a single chip between the individual synapses when used without calibration. 

Note that this spread of transfer functions due to the presynaptic mismatch is not necessarily detrimental; it could be exploited in the context of e.g. liquid computing \cite{maass02} or in the Neural Engineering Framework (NEF) \cite{eliasmith03}, which both rely on random projections via synaptic and neuronal mismatch. However, both need well-controlled readout weights to collapse the random projections. Thus, the 4~bit weight resolution as shown in Fig. \ref{fig_synapse_transfer} together with the ability to set each synapse excitatory or inhibitory could be applied in the NEF to sophisticated population-based signal processing \cite{mayr14b}.\subsection{Overall Results}

The characterization results reported in the previous sections show that all components of the system, such as presynaptic adaptation, synapses and neurons are fully functional. Table \ref{tab_comparison} gives a comparison with state-of-the-art conventional neuromorphic systems and those targeted at biological interfaces.

Using the mixed-signal SC approach, we could aggressively scale down the neuromorphic system, taking full advantage of technology shrink. As synapse area is a major determinant of overall system size for neuromorphic systems \cite{hasler13}, we have included synapse area in the comparison. As expected, our implementation exhibits full technology shrink when compared with for example the synapse area of \cite{indiveri06}.

Conventional neuromorphic systems based on subthreshold circuits \cite{indiveri06} usually do not scale that well, as transistors need to be a certain minimum size to control mismatch \cite{PinedadeGyvez2004,Kinget2005}. There are efforts to overcome this barrier by implementing synapses using analog floating gate storage \cite{basu10}, which is largely immune to mismatch. It could be worthwhile to explore this approach in advanced technology nodes, as floating gates continue to be scaled. However, it is not clear whether the precise storage of analog values required for this approach scales to deep submicron technologies. Current examples of this technique are still implemented in nodes around 350~nm \cite{brink13}, so absolute synapse sizes are still a factor of 10 larger than in our implementation \cite{hasler13}. The neuromorphic system of \cite{seo12} in 45~nm only contains externally-programmable 1-bit synapses in the same overall area and power budget. Thus, even compared to a purely digital neuromorphic system in deep-submicron, our SC system delivers the same or better computational density at a competitive power consumption, see table \ref{tab_comparison}.

\begin{figure}
\centering
\includegraphics[width=0.5\textwidth]{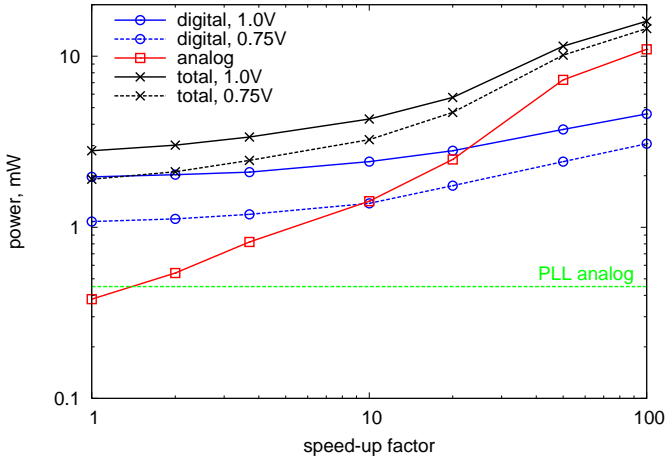}
\caption{\label{fig_power} Power measurements for the different supply voltage domains versus the employed speed-up factor. The central clock divider was set inversely proportional to the speed-up factor, such that the resulting clock frequency scaled linearly with the speed-up (see Sec. \ref{sec_design_digital}). The analog power draw is dominated by the OpAmps (see Sec. \ref{sec_design_neuron}). Their external bias current was chosen such that the system was still operational at the selected speed-up factor. The power consumption is largely independent of the input or output spike rates.}
\end{figure}

As shown in Fig. \ref{fig_power}, the power consumption of the digital circuit parts dominates overall power draw in real-time operation (speed-up factor 1 in the diagram).
Note that the design was not primarily optimized for low power, meaning that all the digital components of the whole IC (not just the neuromorphic system) are permanently connected to the digital supply voltage (with only the clock of the neuromorphic system being switched on), which results in a static power draw of approximately 1.1mW at nominal supply voltage of 1~V.
Furthermore, the design was not optimized for aggressive supply voltage scaling, as done in \cite{seo12}.
Therefore, the lowest digital supply voltage where the digital parts operate without errors is 0.75~V.
We thus performed measurements both at 0.75V and at the nominal supply voltage of 1.0~V.
As expected, the lower supply voltage reduces digital power draw by almost a factor of two.
As described in Sec. \ref{sec_design_system}, our neuromorphic system can be scaled in speed by varying the clock frequency, so that experiments can be performed either in realtime for interfacing to real-world sensors, or in accelerated time for reducing simulation time.
When moving to accelerated simulations, contributions to the power budget change, as can be seen in Fig. \ref{fig_power}.
Both dynamic digital and analog power increase approximately linear with the speed-up factor, the latter because of the increasing required bandwidth of the opamps in the analog part.
As a consequence, analog power dominates for high speed-up factors.
This power draw could be reduced by switching off the opamps during switching phases when their outputs are not used.
The static power draw of the PLL could be reduced if instead of the current fixed-frequency PLL and subsequent clock divider, a variable-frequency PLL such as that in \cite{eisenreich09} was used, where power consumption scales with the output clock frequency.

If power consumption is normalized with respect to the speed-up of the simulation, effective power consumption reduces from 1.9~mW in realtime operation to approximately 15~mW/100$=$0.15~mW for an speed-up factor of 100.
In other words, the energy required for emulating a spiking neural network for one second reduces from 1.9~mJ to 0.15~mJ.
This is mainly due to reduced influence of static power.
Thus, accelerated simulations could be used for increasing energy efficiency for applications that do not require real-time operation.
When assuming that all neurons fire with their maximum frequency of 1~kHz in real-time operation (resulting in 100~kHz at a speed-up factor of 100), the above values correspond to 30~nJ/spike in real time operation and 2.3~nJ/spike at an speed-up factor of 100. This number is well within the range otherwise reported for power-optimized subthreshold architectures, see table \ref{tab_comparison}. The value given for \cite{seo12} counts only the incremental increase in power consumption per additional spike. If the metric of our system (overall power consumption divided by cumulative spike rate) is applied, the energy per spike would be 19~nJ \cite{merolla12}.

In contrast to a fully addressable synaptic matrix \cite{indiveri06}, our architecture inherently relies on feeding all synapses of a given row with the same presynaptic input.
In this way, all components of the synapse circuit that depend solely on the input spikes are shared between synapses, thus they are implemented only once per row, which greatly reduces overall circuit area compared to a fully adressable synaptic matrix.
Memristive arrays \cite{alibart12,mayr12b} inherently use the same architecture, as the employed implementation as a crossbar does not allow for individual presynaptic circuits inside the synaptic matrix.

Driving all the synapses of a row with the same presynaptic input poses constraints on the realizable connection topologies.
Networks that employ all-to-all connectivity or similar topologies with high local connection density can be realized efficiently, whereas for topologies with low connection density, only a fraction of the synapses in the matrix are used.
Dedicated mapping algorithms can partially compensate for these restrictions \cite{galluppi12,mayr07a}.
Increasing the number of presynaptic circuits and letting individual synapses choose between several (e.g. two) inputs also greatly reduces the imposed constraints and makes the architecture well-suited even for topologies with low connection density, as demonstrated in \cite{noack10}.
At the same time, this concept retains the original approach of shared presynaptic circuits, so that the implementation presented in this paper could be easily extended in this way.

In terms of interfacing to biological tissue, our approach is similar to \cite{vogelstein08}, i.e. it concentrates on the behavioural dynamics, while using conventional lab equipment to detect and record biological spikes and convert spikes of the SC neurons back into stimulation signals. 
A reasonable level of versimilitude in the reproduction of physiological behaviour is needed in an interface to neural tissue \cite{vogelstein08}. The chosen short term plasticity has a firm grounding in biological measurements \cite{markram98}. The long term plasticity rule chosen for this implementation has a more theoretical background, with only limited support from biological evidence \cite{brader07}. 
However, our SC implementation is by no means restricted to this single plasticity rule. In particular, the faithful reproduction of pre- and postsynaptic waveforms (evident for example in Fig. \ref{fig_presyn_facil_comb} and Fig. \ref{fig_wave_psp}) could also be employed by a plasticity rule based on neuronal waveforms such as that in \cite{mayr10b}, which aims at the replication of a wide range of biological plasticity experiments \cite{mayr10a}. 

\begin{table*}[ht]
\caption{\label{tab_comparison}Comparison of the presented neuromorphic neural interface with other general-purpose neuromorphic work (upper part) and neuromorphic circuits targeted at biological interfaces (lower part).}
\newcommand{\abcd}{p{1.7cm}}
\centering
\begin{threeparttable}
        \begin{tabular}{|p{1.7cm}||p{0.8cm}|p{1.1cm}|p{1.1cm}|p{1.1cm}|p{1.0cm}|p{1.0cm}|p{1.0cm}|p{1.0cm}|p{1.1cm}|p{3.0cm}|}
             \hline
             Comparison & Ref. & Techn. & System area & Synapse area & Supply voltage & Power & Energy/ spike& Number of input channels & Number of output channels & Special features\\
            \hline
            \multirow{3}{*}{\parbox{0.0cm}}{Conventional neuromorphic systems} & \multicolumn{1}{p{0.7cm}|}{\cite{seo12,merolla12}} & \multicolumn{1}{l|}{45~nm} & \multicolumn{1}{l|}{4.2~mm$^2$} & \multicolumn{1}{l|}{1.6~$\mu$m$^2$} & \multicolumn{1}{l|}{0.53~V} & \multicolumn{1}{l|}{5~mW} & \multicolumn{1}{l|}{45~pJ} &\multicolumn{1}{l|}{--} & \multicolumn{1}{l|}{256} & \multicolumn{1}{p{3.0cm}|}{64k 1-bit synapses, set externally}\\
				   \cline{2-11}
            & \multicolumn{1}{l|}{\cite{indiveri06}} & \multicolumn{1}{l|}{800~nm} & \multicolumn{1}{l|}{1.6~mm$^2$} & \multicolumn{1}{l|}{4495~$\mu$m$^2$} & \multicolumn{1}{l|}{--} & \multicolumn{1}{l|}{1.9~mW\tnote{1}} & \multicolumn{1}{p{0.7cm}|}{900~pJ-1$\mu$J} & \multicolumn{1}{l|}{256} & \multicolumn{1}{l|}{32} & \multicolumn{1}{p{3.0cm}|}{Long- and short term plasticity, neuronal adaptation}\\
				   \hline
              &  This work & 28~nm & 0.36~mm$^2$ &  13~$\mu$m$^2$ &  1~V (analog) 0.75~V (digital) & 1.9~mW\tnote{2} & 2.3nJ-30nJ & 128 &  64 & Short-Term \& Long-Term Plasticity, 8k Synapses for High-Dimensional Closed Loop Processing\\
            \hline
            \multirow{1}{*}{\parbox{0.0cm}}{Biologically-targeted neuromorphic systems} & \multicolumn{1}{l|}{\cite{vogelstein08}} & \multicolumn{1}{l|}{500~nm} & \multicolumn{1}{l|}{7.26~mm$^2$} & \multicolumn{1}{l|}{0.032~mm$^2$} & \multicolumn{1}{l|}{3.3~V} & \multicolumn{1}{l|}{8.3~mW} & \multicolumn{1}{l|}{--} & \multicolumn{1}{l|}{19} & \multicolumn{1}{l|}{10} & \multicolumn{1}{p{3.0cm}|}{Central pattern generator with 190 synapse processing for closed loop BMI}\\
				   \hline            
        \end{tabular}
   \begin{tablenotes}
\item[1] The power for \cite{indiveri06} includes only neuron circuits, not pulse handling and synapses. 
\item[2] Measured power consumption is for the entire system of Fig. 1.
\end{tablenotes}
\end{threeparttable} 
\end{table*}\section{Conclusion}

We have constructed a mixed-signal neuromorphic system implemented in the 28~nm node. The usage of switched capacitor circuit techniques together with dedicated low-leakage optimization allows us to achieve biological-realtime operation. The SC implementation in 28~nm enables a very agressive area scaling without compromising analog performance (see Sec. \ref{sec_results}).
As can be seen from table \ref{tab_comparison}, its power budget is competitive with recent power-optimized digital or analog neuromorphic systems \cite{indiveri06,seo12}. 

In terms of neural recording and stimulation, the high-density system integration in a 28~nm technology, the realistic synaptic and neuronal dynamics and moderate power dissipation make our system a good candidate for future implanted closed loop interfaces (when enhanced by amplifiers and spike detectors). Compared to the biologically targeted neuromorphic system presented in \cite{vogelstein08}, which is optimized for application as a spinal cord central pattern generator, our system contains significantly more neurons and various adaptation mechanisms, which are widely configurable.
In collaboration with the group of S. Marom, one of the target uses of our system is replicating the experiment described in \cite{levy12} with one of the biological networks in the chain replaced by a hardware network. Overall, our system fits very well with this intended usage in the context of biological interfaces. 

However, the presented neuromorphic system is by no means restricted to biological interfaces. The versatility and configurability of the implemented neuron and synapse circuits can be used in general neuromorphic processing comparable to \cite{indiveri10,moradi13}. One interesting use may be in an integrated adaptive vision system \cite{mayr07c} that directly incorporates the high-density neuromorphic processing with deep-submicron pixel cells \cite{henker07}. The configurable-timescale waveform generation shown for example in Fig. \ref{fig_wave_psp} or \ref{fig_presyn_depression} could also be used as a high-density driver for nanoscale memristive arrays \cite{ou13,mayr12b,you14}.

\section*{Acknowledgements}
This work is partly supported by 'Cool Silicon', the 'Center for
Advancing Electronics Dresden' and the European Union 7th framework program, project 'CORONET' (grant no. 269459).

\bibliographystyle{IEEEtran}

\end{document}